\documentclass[runningheads]{llncs}
\usepackage[T1]{fontenc}
\usepackage{comment}
\usepackage{xcolor}
\usepackage{graphicx}
\usepackage{amsmath} 
\usepackage[most]{tcolorbox}
\usepackage{tabularx}
\usepackage{array}
\usepackage{booktabs}
\usepackage{multirow}
\usepackage{tabularx}
\usepackage{array}

\begin{document}

\title{Single and Multi Truth Data Fusion using Large Language Models}
%
%
\author{Hira Beril Kucuk\orcidID{0009-0002-4829-2156} \and
Norman W Paton\orcidID{0000-0003-2008-6617} \and
Jiaoyan Chen\orcidID{0000-0003-4643-6750} \and
Zhenyu Wu \orcidID{0000-0003-0981-5567}}
\authorrunning{H.B. Kucuk et al.}
%
\institute{Department of Computer Science \\ 
University of Manchester,
Oxford Road, Manchester M13 9PL, UK
\email{\{hiraberil.kucuk,norman.paton,jiaoyan.chen,zhenyu.wu\}@manchester.ac.uk}}
\maketitle              
\begin{abstract}

Data fusion, also known as truth discovery, is a data integration problem that aims to determine the correct value or set of values for each attribute of an object when presented with potentially conflicting values from multiple sources. Data fusion tasks belong to two main categories: single-truth scenarios, where each attribute has only one correct value, and multi-truth scenarios, where multiple values can be valid simultaneously.
This paper investigates the use of Large Language Models (LLMs) in data fusion tasks for tabular data. Various prompting strategies, encompassing both single-truth and multi-truth scenarios, are investigated empirically.  Domain-dependent, domain-independent, zero-shot and one-shot prompts are evaluated on three different benchmark datasets. Experimental results demonstrate that LLM-based approaches outperform traditional unsupervised truth discovery methods, such as DART and LTM, across all datasets. The codebase of this study has been made publicly available on GitHub\footnote{\url{https://github.com/hiraberil/LLM-DataFusion}}.

\keywords{Data Fusion, Truth Discovery, Large Language Model}
\end{abstract}

\section{Introduction}

In recent years, the volume, variety and velocity of generated data has been rapidly increasing. This growth has made not only storing data but also obtaining meaningful and reliable information a significant research problem. In this context, data integration has become a prominent research area. Data integration can be characterised as including two main levels namely schema-level integration and instance-level integration \cite{Mandreoli2019DealingAlgorithms}\cite{Putrama2024HeterogeneousOpportunities}.
Schema-level integration, as the initial stage of data integration, structurally and semantically aligns schemas from different sources and creates a common global schema; however, inconsistencies in the underlying data records may not be fully resolved at this stage.  Instance-level data integration aims to integrate actual data records from different data sources. The primary goal of this stage is to identify records that represents the same real-world entity but have been recorded in different formats, incompletely, or inconsistently, and to merge them under a single, consistent representation. 

\begin{figure}[tb!]
\centering
\small
\scalebox{0.7}{%
\begin{minipage}{\textwidth}
\centering
\small

\begin{minipage}[t]{0.39\textwidth}
\begin{tcolorbox}[title=Source 1, colback=white, colframe=black, boxrule=0.6pt]
\textbf{Title:} \textit{The Hobbit}\\
\textbf{ISBN:} \texttt{978-0547928227}\\
\textbf{Author:} J.R.R. Tolkien\\
\textbf{Genre:} Fantasy\\
\textbf{Publication Year:} 1937\\
\textbf{Language:} English
\end{tcolorbox}
\end{minipage}
\hfill
\begin{minipage}[t]{0.57\textwidth}
\begin{tcolorbox}[title=Source 2, colback=white, colframe=black, boxrule=0.6pt]
\textbf{Title:} \textit{The Hobbit: There and Back Again}\\
\textbf{ISBN:} \texttt{9780547928227}\\
\textbf{Author:} John Ronald Reuel Tolkien\\
\textbf{Genre:} Fantasy; Adventure\\
\textbf{Publication Year:} 1937\\
\textbf{Language:} en
\end{tcolorbox}
\end{minipage}

\vspace{0.8em}

\begin{minipage}[t]{0.39\textwidth}
\begin{tcolorbox}[title=Source 3, colback=white, colframe=black, boxrule=0.6pt]
\textbf{Title:} \textit{The Hobbit}\\
\textbf{ISBN:} \texttt{9780547928227}\\
\textbf{Author:} Tolkien, J. R. R.\\
\textbf{Genre:} Children's literature; Fantasy\\
\textbf{Publication Year:} 1937\\
\textbf{Language:} English
\end{tcolorbox}
\end{minipage}
\hfill
\begin{minipage}[t]{0.57\textwidth}
\begin{tcolorbox}[title=Fused record, colback=white, colframe=black, boxrule=0.8pt]
\textbf{Title:} \textit{The Hobbit}\\
\textbf{ISBN:} \texttt{9780547928227}\\
\textbf{Author:} J.R.R. Tolkien\\
\textbf{Genre:} \{Fantasy, Adventure, Children's literature\}\\
\textbf{Publication Year:} 1937\\
\textbf{Language:} English
\end{tcolorbox}
\end{minipage}

\end{minipage}%
}
\caption{An example of data fusion for a book entity. }
\label{fig:book_fusion_boxes}
\end{figure}

Data integration at the instance level typically involves the steps of format transformation, entity resolution and data fusion  \cite{Mandreoli2019DealingAlgorithms}\cite{Tian2025DataChallenges}. 
In an approach composed of these three steps in order, records are first standardised in format, then records representing the same real-world entity are identified, and finally, they are merged to create a single, consistent record \cite{Bleiholder2009DataFusion}\cite{Tian2025DataChallenges}.
Consider the following example: two different sources report the ISBN of a book as ``\textit{ISBN 9780547928227}'' and ``\textit{978-0547928227}''. 
Format transformation may represent both using the format ``\textit{9780547928227}'', entity resolution will group different records containing equivalent ISBNs as representing the same book, and data fusion merges the book records into a single representation, as illustrated in Figure \ref{fig:book_fusion_boxes}.

Consider another example of fusion in the multi-truth setting, where multiple values of an attribute may be correct: three different sellers may report the genre of a book as being ``\textit{Fantasy}'', ``\textit{Fantasy; Adventure}'', and ``\textit{Children’s literature; Fantasy}''. The above data integration steps may output a record with all the valid genre values ``\textit{Fantasy, Adventure, Children’s literature}'', as illustrated in Figure \ref{fig:book_fusion_boxes}. 

Data fusion methods face a number of challenges, particularly in multi-truth scenarios~\cite{Bleiholder2009DataFusion,DBLP:journals/tbd/WangZSLSCZYG25}. A significant portion of existing methods make an assumption that attributes provide a single true value, and are therefore inadequate for modeling situations where multiple true values can exist \cite{10.1145/2983323.2983767}. 
While current unsupervised methods offer partial solutions using evidence from the data provided, they are mostly based on fixed assumptions and may fail to adequately account for context-sensitive semantic differences. Furthermore, the representation of the same real-world entity in different forms (e.g., ``John Ronald Reuel Tolkien'' and ``J.R.R. Tolkien'') requires normalisation and semantic matching, a process not explicitly addressed in most traditional methods.
LLMs are capable of modeling semantic relationships in natural language and making contextual inferences \cite{wei2022chain}. These capabilities offer potential advantages in addressing challenges such as implicit normalisation, expression diversity, and the interpretation of contradictory information encountered in data fusion. In particular, tasks such as comparing heterogeneous values from different sources at the semantic level and evaluating both single and multi-truth values together can be handled more flexibly with the context-based inference capabilities of LLMs. In this study, the data fusion problem is addressed as an LLM-supported data integration process.

The main contributions of this paper are as follows:
\begin{itemize}
    \item An exploration of the use of LLMs for data fusion that includes zero-shot, single-shot, domain-independent and domain-dependent prompts for both single-truth and multi-truth settings.
    \item An empirical evaluation in various benchmark scenarios in comparison with four classic data fusion methods.
    \item An analysis of the strengths, limitations and reliability of LLM-supported data fusion.
\end{itemize}

The remainder of this paper is structured as follows. Section 2 reviews related work on data fusion and LLM applications in data integration. Section 3 presents prompt-based strategies for data fusion and details the proposed methodology. Section 4 describes the experimental setup and evaluation criteria, and presents and analyses the results. Section 5 concludes the paper.

\section{Related Work}

\subsection{Data Fusion}

Data fusion is a critical step in instance-level data integration.
It aims to reconcile potentially conflicting records from different sources, which are often obtained after schema matching, entity resolution, and format transformation, to achieve a more consistent and reliable record.  
{Figure~\ref{fig:book_fusion_boxes} illustrates three records of a book entity that need to be fused. For single-truth attributes such as ISBN, fusion is expected to output one correct value, whereas for multi-truth attributes such as author and genre, fusion is expected to output one or more correct values.}

{Bleiholder and Naumann classify strategies for handling conflicting values into conflict-ignoring, conflict-avoiding and conflict-resolving approaches \cite{Bleiholder2009DataFusion}.
In practice, however, most research attention has focused on conflict-resolving methods, particularly truth discovery, which aims to infer the most credible value from multiple conflicting observations by jointly estimating source reliability and value correctness \cite{Dong2150LessIntegration}. As summarised in more recent surveys, truth discovery methods are commonly characterised by how they model observations, source dependencies, and learning settings, and many unsupervised methods follow either a voting-style aggregation scheme or an iterative/probabilistic estimation process \cite{LiTruthSolved,DBLP:journals/tbd/WangZSLSCZYG25}.} Majority voting is one of the most straightforward approaches, which selects the value supported by the largest number of sources. More advanced truth discovery methods go beyond this strategy by jointly modelling factors such as source reliability and source dependence. 
Many such methods follow an iterative estimation paradigm, with detailed taxonomies and comparisons in recent surveys (e.g., \cite{10.1145/2897350.2897352,DBLP:journals/tbd/WangZSLSCZYG25}) complementing  descriptions of individual proposals (e.g., \cite{Dong2009IntegratingDependence,DongDataIntegration,LiTruthSolved}).

Here we briefly describe representative proposals that are used later as baselines in the experiments. The Latent Truth Model (LTM) adopts a probabilistic graphical model-based approach, treating the truth of each claimed value as a latent variable and jointly modelling it with source reliability \cite{Zhao2150AIntegration}. It supports the multi-truth setting.
Domain-Aware Truth Discovery (DART) extends classical bayesian approaches by considering expertise of sources in different domains \cite{PvldbReferenceFormatXueling2018Domain-AwareSources}. While traditional methods generally assign a single reliability score to a source, DART calculates domain-based reliability scores for each source. This incorporates the fact that a source may be reliable in one domain but weak in another. These scores are used to calculate the probability of accuracy of the values. 

\subsection{LLMs in Data Integration}

Recent studies demonstrate how LLMs can contribute to data integration, and proposals have been made that investigate how LLMs can support many different data integration tasks. These include schema matching~\cite{LiuPSWF25}, relationship discovery~\cite{Dong0NEO23}, entity resolution~\cite{ZeakisPSK25}, format transformation~\cite{NobariR25} and schema inference~\cite{DBLP:conf/esws/WuCP25}. As well as these proposals that focus on individual steps, several papers have explored how LLMs can contribute to data integration pipelines (e.g., \cite{10.1007/978-3-032-02215-8_2,QianHZHMWWSLDZ24,steiner2026automaticendtoenddataintegration}). These proposals may use embeddings to support semantic similarity or generative techniques to construct integrated representations at the schema or instance level.

To the best of our knowledge, the following two studies represent the only existing LLM-based approaches that relate to data fusion. Both are part of multi-step pipelines.
Ji \textit{et al}. \cite{Ji2025TableResolution} discusses table integration in data lakes by dividing it into three sub-tasks: pairwise integrability judgment, integrable set discovery, and multi-tuple conflict resolution, where the latter step corresponds to data fusion.  A single prompt is proposed that can be considered to be domain-dependent, multi-shot and single-truth. This prompt is not compared with other prompts or baseline data fusion methods.  Thus we complement and go further than this work in terms of data fusion by considering a wider range of prompt types, baselines and datasets. 

Steiner and Bizer \cite{steiner2026automaticendtoenddataintegration} explore automating the end-to-end data integration process, encompassing schema matching, value normalization, entity matching, and data fusion, through LLMs. The proposal utilizes methods from an existing Python Data Integration library (PyDI), and LLMs play different roles at different stages in the pipeline. In terms of data fusion, the LLM is positioned as a tool that guides the configuration of existing methods rather than directly making the final data merging decision.  Thus we complement this work by exploring a different role for the LLM in the data fusion process.

Our work positions LLMs as a truth-discovery component that directly resolves conflicting values from multiple sources without relying on external tools, and evaluates this approach on widely used benchmark datasets. %

\section{LLM Prompts for Data Fusion}

The LLM prompts used in this study for data fusion can be classified into four types according to whether they are domain-independent (DI) or domain-dependent (DD), and whether they are for single-truth (ST) or multi-truth (MT). These prompt categories are denoted as DI-ST, DI-MT, DD-ST and DD-MT.

DI prompts can be applied to any domain without modification, regardless of the type of data to be combined; abstract terms such as entity, source and value are used. DD prompts, on the other hand, directly include terms associated with the target attribute (author, director, departure time, etc.) and need to be revised for a different domain.
ST prompts are structured for scenarios where only one true value exists for an attribute (e.g., departure time or gate), while MT prompts are structured for scenarios where multiple valid values are possible (e.g., authors or directors).

\begin{figure}[tb!]
    \centering
    \includegraphics[width=1\linewidth]{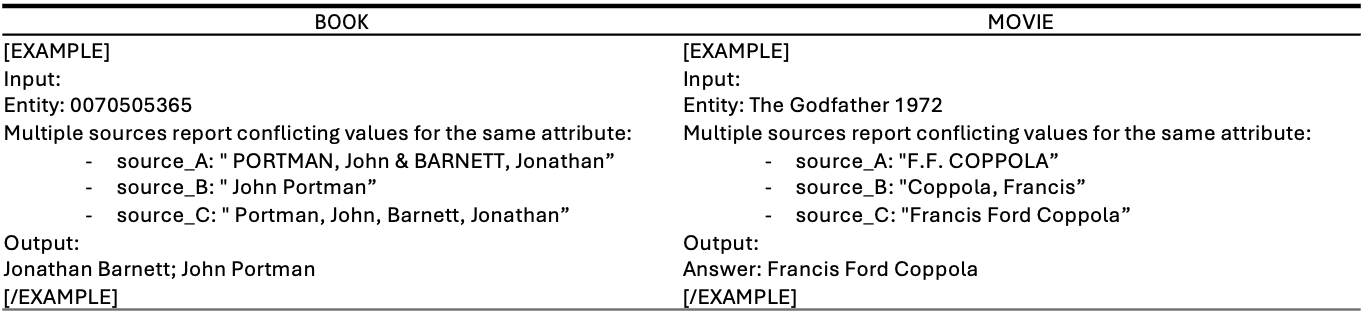}
    \caption{Domain-Independent Examples – Multi-Truth Book \& Movie Datasets. Domain-Dependent Examples are identical except that the \textit{Entity} is referred to using its domain-specific type such as \textit{Book ISBN} or \textit{Movie}, and the \textit{sources} are labelled as \textit{sellers} in the Book dataset and as \textit{sources} in the Movie dataset.}
    \label{fig:multi-example}
\end{figure}

\begin{figure}[tb!]
    \centering
    \includegraphics[width=0.5\linewidth]{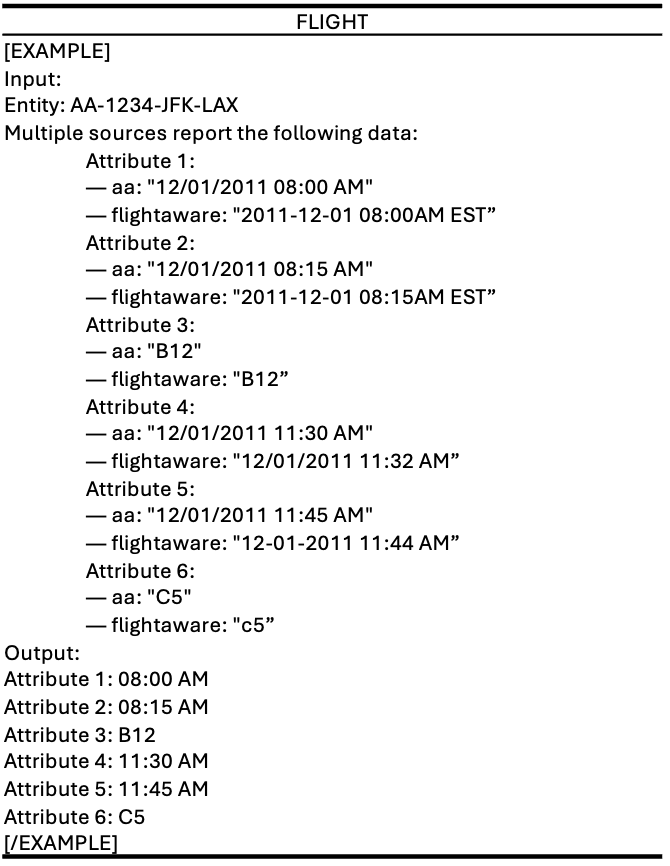}
    \caption{Domain-Independent Examples – Single-Truth Flight Dataset. Domain-Dependent Examples are identical except that the \textit{Entity} is referred to as \textit{Flight ID},  and the generic attribute labels such as \textit{Attribute 1} are replaced by their domain-specific names such as \textit{Scheduled departure}.}
    \label{fig:single-example}
\end{figure}

Book, movie and flights datasets are used in the experiments reported in Section \ref{sec:evaluation}.
As the book and movie datasets used for evaluation in this study contain multi-valued attributes, the DI-MT and DD-MT prompts were applied to these datasets. Since the flight dataset consists of six single-valued attributes, the DD-ST and DI-ST categories were used for this dataset. 
Furthermore, each prompt was evaluated under both zero-sample (0-shot) and one-sample (1-shot) settings. In the 1-shot setting, a fixed input-output pair, as presented in Figures \ref{fig:multi-example} and \ref{fig:single-example}, is added to the beginning of the prompt; in the 0-shot setting, there is no such example block.

\begin{figure}
    \centering
    \includegraphics[width=1\linewidth]{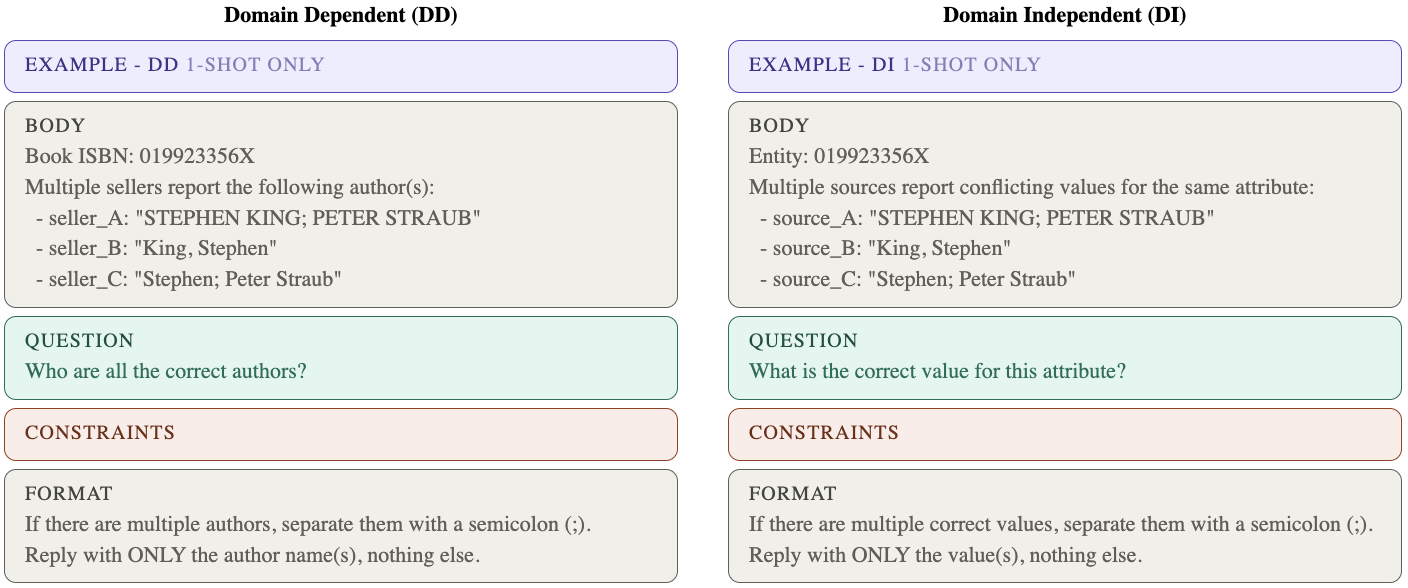}
    \caption{Design of Multi-Valued Prompt Structures for the Book Dataset. The prompt structure for the Movie dataset is identical, with the entity referred to as \textit{Movie: {title - year}} instead of \textit{Book ISBN: {isbn}}.}
    \label{fig:MVprompts}
\end{figure}

\begin{figure}
    \centering
    \includegraphics[width=1\linewidth]{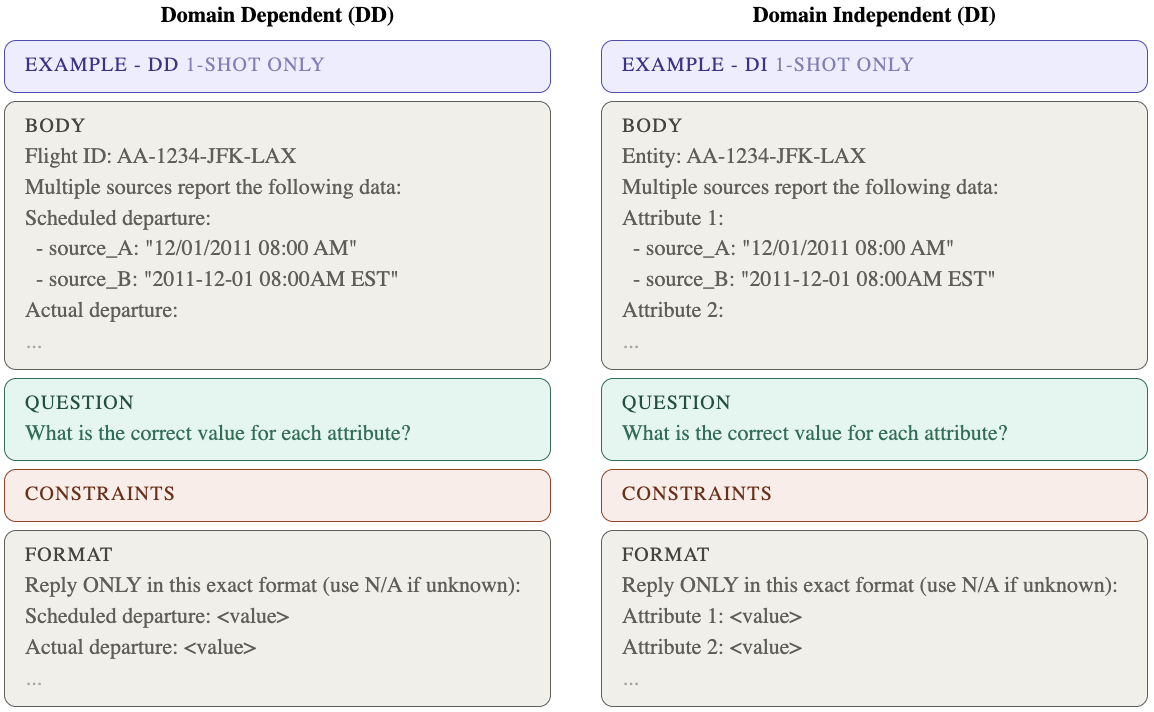}
    \caption{Design of Single-Valued Prompt Structures for the Flight Dataset}
    \label{fig:SVprompts}
\end{figure}

The overall prompt structures are presented in Figures \ref{fig:MVprompts} and \ref{fig:SVprompts}. Each prompt consists of an optional EXAMPLE block, a BODY block presenting the table data, a QUESTION block containing the task instructions, an optional CONSTRAINTS block containing additional constraints that modify the prompt's behaviour, and a FORMAT block defining the expected output format. 
The prompts presented in these two figures omit the details of the CONSTRAINTS block; they are presented as follows.

\paragraph{Prompt Constraints.}

Two constraints may be included in prompts, specifically:

\begin{itemize}
\item \textit{C1: Only use values that appear in the sources above}. This instructs the LLM not to invent values, but instead to choose from the values provided.  

\item \textit{C2: If the same value appears in different formats, count them as one}. This instructs the LLM to overlook representational inconsistencies.
\end{itemize}

\section{Experimental Evaluation}
\label{sec:evaluation}

\subsection{Experimental Setup}

\paragraph{Baselines.} 

The following truth discovery methods are used as the baselines.
(1) \textbf{Majority Vote (MV)} chooses the value with the highest frequency among all the input values for each attribute. This is a simple method without considering source reliability but involves the least computation. According to \cite{Dong2150LessIntegration}, its performance decreases in the multi-truth setting.
(2) \textbf{Source Reliability Vote (SRV)} weights the voting process by iteratively estimating source reliability \cite{Li2014ResolvingEstimation}. In this framework, values obtained from sources with higher reliability receive more weight; source reliability and value accuracy are updated iteratively. It can be regarded as a weighted extension of the basic voting approach.
(3) \textbf{LTM} is a Bayesian framework that models source reliability and value accuracy as latent variables \cite{Zhao2150AIntegration}. 
(4) \textbf{DART} is a domain-aware algorithm based on the observation that the reliability of the same data source can differ significantly across different domains \cite{PvldbReferenceFormatXueling2018Domain-AwareSources}. Domain expertise, value accuracy, and source reliability are updated concurrently in an iterative Bayesian framework.
The implementations of these baselines are from an open-source Python repository for data fusion\footnote{\url{https://github.com/yishangru/TruthDiscovery}}, and the default parameters from the original papers were used.

\paragraph{Benchmark Datasets.}

The following criteria were considered in selecting the datasets: representation of different domains,  coverage of both single- and multi-truth settings, high density of inter-source discrepancies, and availability of the ground truth.
Based on these criteria, the following three real-world datasets were used for evaluation,
where two are multi-truth and one is single-truth. 
(1) The \textbf{\textit{Book}} dataset is obtained from an open-source repository \cite{Yin2007Truth}\cite{Dong2009IntegratingDependence}\cite{lunadong_datasets2026}. It contains author lists from AbeBooks.com, across 894 sources. The multi-valued nature of the author lists positions the dataset as a benchmark with multiple truths.
(2) The \textbf{\textit{Movie}} dataset is derived from an open-source data repository \cite{heathersherry2026}\cite{PvldbReferenceFormatXueling2018Domain-AwareSources}. It contains director and year information for films compiled from 15 different web sources, including IMDb, Amazon and Metacritic. Its high levels of source contradiction and multi-accuracy structure provide a second basis for multi-truth comparison. 
(3) The \textbf{\textit{Flight}} dataset is taken from the same open-source repository as the book dataset \cite{lunadong_datasets2026}\cite{LiTruthSolved}. It includes departure and arrival information for over 1,200 flights compiled from 38 different sources. Since there is only one correct value for each object, it provides a single-truth scenario. 

\paragraph{Large Language Models (LLMs).}

Language understanding capacity, cost and API accessibility have been considered in selecting LLMs for evaluation.
In the experiments, the GPT-4o-mini model was primarily preferred. It was observed that GPT-4o-mini has sufficient language understanding capacity for parsing tasks of this scale and was found to be suitable in terms of cost-performance balance. Unless otherwise specified, the results in this paper are based on GPT-4o-mini.
To evaluate the impact of model selection on performance, the prompt configuration that provided the highest performance for each dataset was also run with the GPT-4o and Claude Sonnet 4 models.
For LLM runs, the temperature was set to 0 and the maximum number of tokens was set to 256.

\paragraph{Evaluation Metrics.}

All the methods were evaluated using Recall, Precision, and F1-score, which are widely used in the data fusion literature \cite{PvldbReferenceFormatXueling2018Domain-AwareSources} \cite{Dong2150LessIntegration}. The definitions of true positive (TP), false negative (FN), and false positive (FP) differ in single-truth and multi-truth settings, as described below.

Both Ground Truth and predicted values are subject to normalisation. Normalisation involves Unicode NFKD parsing, lowercase conversion, removal of titles from person names, and replacement of punctuation marks with spaces. Model performance is evaluated using the standard confusion matrix definitions.

In the multi-truth experiments, each predicted value is independently evaluated against the set of true values, in accordance with the evaluation protocol proposed in  \cite{Dong2150LessIntegration}. For an entity, where $T = {t_1, t_2, \ldots}$ represents the set of true values and $P = {p_1, p_2, \ldots}$ represents the set of predicted values, a predicted value of $p_j$ is considered to match a true value of $t_i$ only when the word sets (after normalisation) are identical:
\[\text{match}(t_i, p_j) = 1 \iff \text{words}(t_i) = \text{words}(p_j)\]

Let $P$ denote the predicted values, and $T$ denote the values in the Ground Truth (GT). Under this formulation, a value is a true positive (TP) if it is predicted and matches a value in the GT ($P \cap T$), a false positive (FP) if it is predicted but does not match a value in the GT ($P \setminus T$), and a false negative (FN) if it is in the GT but is not predicted ($T \setminus P$). TP, FP and FN counts are then obtained for each fused record. The overall recall, precision and F1 scores are calculated based on the TP, FN and FP counts for each of records:

\[R = \frac{\sum \text{TP}}{\sum \text{TP} + \sum \text{FN}}, \quad P = \frac{\sum \text{TP}}{\sum \text{TP} + \sum \text{FP}}, \quad F1 = \frac{2 \cdot R \cdot P}{R + P}\]

This definition allows classic single-valued baseline methods (DART, LTM, MV, SRV) to be evaluated within the same framework as LLM outputs across both single-valued and multi-valued datasets.

\subsection{Results}

In this section, experimental results are presented for each of the datasets in turn, comparing the baseline methods with DD and DI LLM prompts. 

\subsubsection{Book}

The results on the Book dataset are presented in the left hand side of Table \ref{tab:combined-results}. The results in bold are those associated with the highest F1-score within a group of methods (baselines, domain-dependent and domain-independent).

We can observe that the LLM with DD prompts demonstrate significantly better performance than the baselines and the LLM with DI prompts. The DD models produce reasonably balanced results in terms of recall and precision, and this is reflected in the F1 scores. The fact that the highest F1-scores are achieved in the DD-1shot, DD-C2-1shot, and DD-C1C2-1shot configurations demonstrates that the domain-dependent 1-shot prompts are the most effective on this dataset. In contrast, the performance of LLM with DI prompts is significantly lower; particularly in 0-shot scenarios, the low recall and precision values indicate that this approach fails to adapt sufficiently to the dataset. 

When examining the effect of example provision, it is observed that 1-shot generally improves performance over 0-shot. This improvement is particularly pronounced in DI prompts; that is, providing contextual examples offers a critical improvement for models with poor performance. In DD prompts, however, the already high performance increases in a more limited manner.
Regarding the effects of the constraints C1 and C2, it is observed that C2 generally improves performance, whereas C1, when used alone, can sometimes reduce performance. However, the combined use of C1 and C2 is demonstrated to be more effective, particularly in the 1-shot scenario, where it is one of the configurations achieving the highest F1-scores.

An analysis of the baseline methods indicates that the LTM method produces the most balanced results; however, these methods generally prioritise either recall or precision. For example, while DART produces high recall but low precision, SRV has very high precision but quite low recall. This imbalance negatively affects F1 scores. 

Overall, it is concluded that: (1) the DD prompt-based LLM offers more balanced precision and recall, and achieves higher performance compared to the baselines and DI prompt-based LLM; (2) example demonstration enhances LLM performance; and (3) the best result is obtained without the constraints C1 and/or C2.

\subsubsection{Movie}

The results on the Movie dataset are presented in the right hand side of Table \ref{tab:combined-results}. 
When examining the results for the movie dataset, it is observed that both baseline methods and LLM-based approaches exhibit higher and more balanced performance compared to the book dataset. Particularly noteworthy is the significantly better results produced by the LTM (F1=0.7959) and DART (F1=0.7768) methods compared to the book dataset. 
In contrast, the MV and SRV methods are insufficient in this dataset due to their single-value selection bias.

\begin{table*}[tb!]
\centering
\caption{Experimental Results on Book and Movie Datasets. Note 1sh and 0sh denote 1-shot and 0-shot example demonstration, respectively.}
\label{tab:combined-results}
\begin{tabular}{lcccccc}
\toprule
\multirow{2}{*}{Method} 
& \multicolumn{3}{c}{Book Dataset} 
& \multicolumn{3}{c}{Movie Dataset} \\
\cmidrule(lr){2-4} \cmidrule(lr){5-7}
& Recall & Precision & F1 
& Recall & Precision & F1 \\
\midrule

DART & 0.8112 & 0.4758 & 0.5997 & 0.8776 & 0.6967 & 0.7768 \\
LTM 
& \textbf{0.6782} & \textbf{0.7137} & \textbf{0.6955} 
& \textbf{0.8209} & \textbf{0.7725} & \textbf{0.7959} \\
MV & 0.4714 & 0.7423 & 0.5765 & 0.4687 & 0.7696 & 0.5826 \\
SRV & 0.1463 & 0.8570 & 0.2498 & 0.2537 & 0.8763 & 0.3935 \\

\midrule

DD-0shot 
& 0.8339 & 0.7257 & 0.7760 
& 0.8777 & 0.7500 & 0.8089 \\

DD-1shot
& \textbf{0.8102} & \textbf{0.7551} & \textbf{0.7817} 
& \textbf{0.8690} & \textbf{0.7713} & \textbf{0.8172} \\

DD-C1-0shot 
& 0.8237 & 0.6506 & 0.7270 
& 0.8603 & 0.7577 & 0.8057 \\

DD-C1-1shot 
& 0.8051 & 0.7121 & 0.7558 
& 0.8210 & 0.7673 & 0.7932 \\

DD-C2-0shot 
& 0.8356 & 0.7114 & 0.7685 
& 0.8603 & 0.7665 & 0.8107 \\

DD-C2-1shot 
& 0.8237 & 0.7341 & 0.7764 
& 0.8428 & 0.7814 & 0.8109 \\

DD-C1C2-0shot 
& 0.8339 & 0.6703 & 0.7432 
& 0.8734 & 0.7576 & 0.8114 \\

DD-C1C2-1shot 
& 0.8305 & 0.7335 & 0.7790 
& 0.8384 & 0.7773 & 0.8067 \\

\midrule

DI-0shot 
& 0.4627 & 0.3212 & 0.3792 
& 0.8210 & 0.7373 & 0.7769 \\

DI-1shot 
& 0.4746 & 0.4314 & 0.4520 
& 0.7686 & 0.7213 & 0.7442 \\

DI-C1-0shot 
& 0.4780 & 0.3099 & 0.3760 
& 0.8035 & 0.7244 & 0.7619 \\

DI-C1-1shot 
& 0.4034 & 0.3934 & 0.3983 
& 0.7293 & 0.7389 & 0.7341 \\

DI-C2-0shot 
& 0.4729 & 0.3407 & 0.3960 
& 0.8166 & 0.7276 & 0.7695 \\

DI-C2-1shot 
& \textbf{0.5153} & \textbf{0.5477} & \textbf{0.5310} 
& 0.7817 & 0.7490 & 0.7650 \\

DI-C1C2-0shot
& 0.4644 & 0.3258 & 0.3829 
& \textbf{0.8428} & \textbf{0.7366} & \textbf{0.7862} \\

DI-C1C2-1shot 
& 0.4898 & 0.4857 & 0.4878 
& 0.7773 & 0.7479 & 0.7623 \\

\bottomrule
\end{tabular}

\end{table*}

In the examination of LLM-based methods, it is seen that DD prompts  offer the highest and most balanced performance. DD-1shot yields the best result, while DD-C2 and DD-C1C2 variants exhibit similar performance. This demonstrates the strength of the DD prompts. DI prompts, however, have become more competitive in this dataset compared to the book dataset. In particular, the F1-scores of DI-0shot and DI-C1C2-0shot, which approach 0.78, indicate a significant improvement.
 
While performance decreases with DI prompts compared to DD prompts, it appears that DI approaches can also perform well when sufficient semantic signals are provided. Indeed, the fact that the F1 difference between DD and DI methods in the movie dataset remains at a low level of approximately 0.03 shows that the model's prior knowledge is similarly effective in both approaches in this dataset. 

Overall, the performance of both traditional baselines and the LLM-based methods is stronger in the movie dataset; however, the highest and most balanced results are still obtained by LLM with DD prompts. Nonetheless, it is clear that DI prompts have also become quite competitive.

\subsubsection{Flight}

The results on the Flight dataset are presented in the left hand side of Table \ref{tab:flight-combined}.
In the Flight dataset, baseline methods exhibit significantly higher performance compared to the other two datasets. The single-truth problem structure eliminates the need to model multi-truth sets, which has resulted in stronger performance for the single-truth baselines MV and SRV.
However, the fact that different sources report identical time information in different ways due to time zone differences, AM/PM representation, and the presence/absence of date prefixes creates a structural constraint for surface-level matching-based methods and determines the performance ceiling that traditional methods can achieve.

One of the findings is that the DD and DI prompts exhibit similar performance (F1=0.9119 and F1=0.9118, respectively) in this dataset.

\begin{table*}[t]
\centering
\caption{Results on Flight Dataset under Flight ID and Obfuscated ID Settings}
\label{tab:flight-combined}

\begin{tabular}{lcccccc}
\toprule
\multirow{2}{*}{Method}  
& \multicolumn{3}{c}{Flight ID} 
& \multicolumn{3}{c}{Obfuscated ID} \\
\cmidrule(lr){2-4} \cmidrule(lr){5-7}
& Recall & Precision & F1 
& Recall & Precision & F1 \\
\midrule

DART 
& 0.8487 & 0.7667 & 0.8056 
& - & - & - \\

LTM
& \textbf{0.8653} & \textbf{0.7817} & \textbf{0.8214} 
& - & - & - \\

MV 
& 0.8506 & 0.7317 & 0.7867 
& - & - & - \\

SRV 
& \textbf{0.8653} & \textbf{0.7817} & \textbf{0.8214} 
& - & - & - \\

\midrule

DD-0shot 
& 0.9507 & 0.8742 & 0.9108 
& 0.9398 & 0.8626 & 0.8996 \\

DD-1shot
& 0.9489 & 0.8725 & 0.9091 
& \textbf{0.9453} & \textbf{0.8677} & \textbf{0.9048} \\

DD-C1-0shot
&\textbf{ 0.9544} & \textbf{0.8731} &\textbf{ 0.9119 }
& 0.9380 & 0.8595 & 0.8970 \\

DD-C1-1shot
& 0.9398 & 0.8641 & 0.9003 
& 0.9288 & 0.8540 & 0.8899 \\

DD-C2-0shot 
& 0.9215 & 0.8473 & 0.8829 
& 0.9197 & 0.8442 & 0.8803 \\

DD-C2-1shot 
& 0.9361 & 0.8593 & 0.8961 
& 0.9325 & 0.8559 & 0.8926 \\

DD-C1C2-0shot 
& 0.9288 & 0.8526 & 0.8891 
& 0.9361 & 0.8564 & 0.8945 \\

DD-C1C2-1shot 
& 0.9398 & 0.8612 & 0.8988 
& 0.9380 & 0.8595 & 0.8970 \\

\midrule

DI-0shot 
& 0.9380 & 0.8610 & 0.8978 
& 0.8960 & 0.8238 & 0.8584 \\

DI-1shot 
& 0.9161 & 0.8423 & 0.8776 
& 0.9124 & 0.8375 & 0.8734 \\

DI-C1-0shot 
& 0.9106 & 0.8358 & 0.8716 
& 0.8996 & 0.8244 & 0.8604 \\

DI-C1-1shot 
& 0.9434 & 0.8660 & 0.9031 
& 0.9252 & 0.8478 & 0.8848 \\

DI-C2-0shot 
& 0.8777 & 0.8070 & 0.8409 
& 0.8504 & 0.7819 & 0.8147 \\

DI-C2-1shot 
& 0.9142 & 0.8392 & 0.8751 
& 0.8942 & 0.8208 & 0.8559 \\

DI-C1C2-0shot 
& 0.9015 & 0.8261 & 0.8621 
& 0.8577 & 0.7899 & 0.8224 \\

DI-C1C2-1shot
& \textbf{0.9526} & \textbf{0.8744} & \textbf{0.9118}
& \textbf{0.9288} & \textbf{0.8555} & \textbf{0.8906} \\

\bottomrule
\end{tabular}
\end{table*}

\begin{table}[t]
\centering
\caption{Comparison of LLM performance across datasets. Results report the best-performing prompt configuration per dataset for each model.}
\label{tab:model-comparison}

\begin{tabular}{l ccc ccc ccc}
\toprule
\multirow{2}{*}{Dataset}
& \multicolumn{3}{c}{GPT-4o-mini} 
& \multicolumn{3}{c}{GPT-4o} 
& \multicolumn{3}{c}{Claude Sonnet 4.6} \\
\cmidrule(lr){2-4} \cmidrule(lr){5-7} \cmidrule(lr){8-10}
& Recall & Precision & F1 & Recall & Precision & F1 & Recall & Precision & F1 \\
\midrule
Book   & 0.8102 & 0.7551 & 0.7817 & 0.7983 & 0.7572 & 0.7772 & \textbf{0.8254} & \textbf{0.7718} & \textbf{0.7977} \\
Movie  & \textbf{0.8690} & \textbf{0.7713} & \textbf{0.8172} & 0.7118 & 0.7689 & 0.7392 & 0.7991 & 0.8097 & 0.8044 \\
Flight & \textbf{0.9544} & \textbf{0.8731} & \textbf{0.9119} & 0.8467 & 0.7798 & 0.8119 & 0.2153 & 0.7662 & 0.3362 \\
\bottomrule
\end{tabular}%

\end{table}

\begin{table}[b]
\centering
\caption{Financial cost for GPT-4o-mini fusing the benchmark datasets}
\label{tab:cost-comparison}

\begin{tabular}{l r r r r}
\toprule
Dataset & Cost & Input Tokens & Output Tokens & Total Tokens \\
\midrule
Book & \$0.57 & 3.615M & 73,448 & 3.688M \\
Movie & \$0.15 & 0.762M & 26,715 & 0.789M \\
Flight & \$0.48 & 2.811M & 127,181 & 2.938M \\
\bottomrule
\end{tabular}%

\end{table}

With the DD prompts, the source constraint (C1) provides the highest DD performance with F1=0.9119. This constraint discourages the model from generating values not found in the sources, proving relevant for structured time data. 
In contrast, the addition of the format equivalence instruction (C2) reduces performance in both DD and DI prompts; this is most likely because the normalization function already resolves most format differences during the evaluation phase, and C2 creates an additional layer of complexity. With DI prompts, the best result was obtained with the C1C2, 1-shot configuration where all constraints were combined (F1=0.9118).

A detailed look at the DD C1 0 Shot configuration, which yields the best F1 score, shows that results differed from attribute to attribute. 
The planned departure and planned arrival attributes achieved almost perfect accuracy (F1=0.9900 and 1.0000); sources generally agree on these values. In contrast, the actual departure (F1=0.8351) and actual arrival (F1=0.8854) showed lower performance. Error analysis reveals that a large portion of the failures in these attributes stemmed from small discrepancies in the relevant times.

\subsubsection{Obfuscated Flight ID}
In the experiment results reported to date, it is not obvious to what extent the LLM approaches are using background knowledge to select values for inclusion in the fused result. Information on books, movies and flights is publicly available, and can be assumed to be used during LLM training.  Experiments carried out using LLMs on public datasets are open to the criticism that the results may have been significantly different in domains that are less well represented in public data~\cite{BodensohnBVSB25}.

To reduce the ability of the LLM to use background knowledge to fuse flight data, we have repeated the experiments on the flight dataset with an obfuscated Flight ID.  Thus, instead of a Flight Id of the form \textit{Airline-FlightCode-From-To} (e.g., \textit{AA-1007-MIA-PHX}), an Id of the form \textit{FLIGHT-Integer} (e.g., \textit{FLIGHT-001}) has been used, making it difficult for the LLM to use background timetable knowledge during data fusion.
This change was applied only to the LLM-based prompts, because truth discovery methods such as DART, LTM, MV and SRV act only on the values from the column to be fused, and are therefore immune to changes in FlightId. 

The results with the obfuscated Flight Id are presented in the right hand side of Table \ref{tab:flight-combined}.
The change in the FlightId format had a small impact on the F1 score; the best result decreased only slightly, from 0.9119 to 0.9048. Across the range of results, the F1 score dropped in 8 out of 9 DD cases and in 9 out of 9 DI cases, with an overall average reduction of 0.014. This small drop indicates that the LLM derives limited benefit from the semantic information encoded in the original flight IDs. 

Overall, together with the close performance of DD and DI prompts, this suggests that the model’s decisions are driven mainly by the consistency patterns among source claims rather than by domain-specific cues in the entity key.

\subsubsection{LLM Selection}
The results for different LLMs are shown in Table \ref{tab:model-comparison}. Following on from the results obtained with the GPT-4o-mini model, the prompt combinations that yielded the best F1 score were also tested with the GPT-4o and Claude Sonnet 4.6 models. In this approach, whose primary aim was to examine the impact of model selection on performance, Claude Sonnet provided the best F1 score on the Book dataset, but much the worst recall and F1 score with the single-valued prompt on the Flight dataset. The best results were obtained with the GPT-4o-mini in the Movie and Flight datasets.

\subsubsection{Costs}

Table \ref{tab:cost-comparison} presents the costs incurred carrying out data fusion on the benchmark datasets, with the associated dataset sizes. Whether or not these are prohibitive would depend on the application and the user.  The elapsed time per API call for DD-1shot averaged 1.35s/request for Book, 1.14s/request for Movie and 3.28s/request for Flight.

\subsubsection{Summary}
When the results obtained across the benchmark datasets are evaluated as a whole, a number of observations can be made:
\begin{enumerate}
    \item DD LLM-based approaches consistently produce more balanced precision/recall results and higher F1 scores than the baseline methods.
    \item DD prompts provided better F1 scores than the corresponding DI prompts in 8 out of 8 cases in the Book dataset, 8 out of 8 cases in the Movie dataset, and 6 out of 8 cases in the Flights dataset, so we can say that domain-specific wording in prompts is useful for data fusion. We note that the domain-specific wording used should be able to be generated, rather than hand-crafted, in many cases.
    \item The 1-shot prompts outperformed the 0-shot prompts in F1 score in 8 out of 8 cases for Books, 2 out of 8 cases for Movies, and 5 out of 8 cases for Flights, so the impact of examples is inconsistent across the datasets.
    \item The C1 constraint improved the F1 score compared with no constraint in 0 out of 4 cases for the Book dataset, 0 out of 4 cases for the Movie dataset, and 2 out of 4 cases for the Flight dataset.
    The C2 constraint improved the F1 score compared with no constraint in 2 out of 4 cases for the Book dataset, 2 out of 4 cases for the Movie dataset, and 0 out of 4 cases for the Flight dataset.
    The C1+C2 constraints improved the F1 score compared with no constraint in 2 out of 4 cases for the Book dataset, 3 out of 4 cases for the Movie dataset, and 1 out of 4 cases for the Flight dataset. As such, the experiments do not provide compelling evidence for including the constraints.
    \item Obfuscating the Flight ID in the Flights dataset led to a small but consistent reduction in performance. In 15 out of 16 cases, the LLM prompts with the obfuscated Flight ID provided a higher F1 score than the best performing baseline.
\end{enumerate}

\section{Conclusions}

Large language models are having a substantial impact on many data intensive tasks. To the best of our knowledge, this paper provides the first wide-ranging proposal for, and evaluation of, prompt-based approaches to data fusion.

Here we revisit the claimed contributions of the paper from the introduction to highlight the nature of the contributions made:
\begin{enumerate}
\item \textit{An exploration of the use of LLMs for data fusion that includes zero-shot,
single-shot, domain-independent and domain-dependent prompts for both single-truth and multi-truth settings}. Separate prompts have been proposed for single and multi-truth settings, and in both such settings 0-shot and 1-shot, domain-dependent and domain-independent variants have been derived. The addition of constraints to guide the LLM in how to perform data fusion has also been investigated.
\item \textit{An empirical evaluation in various benchmark scenarios in comparison with
four classic data fusion methods}. The various prompts from (1) have been compared with 4 (also unsupervised) baseline proposals on 3 benchmark datasets.  The baselines include majority voting and three high profile techniques published in SIGMOD and PVLDB that between them have well over 1000 Google Scholar citations. The domain-dependent LLM-based proposals consistently outperformed all the baselines on all the datasets.
\item \textit{An analysis of the strengths, limitations and reliability of LLM-supported data fusion}. The wide-ranging evaluation provides various insights into prompt design for data fusion. Including domain-specific terminology in prompts sometimes provided significant benefits, but the provision of examples and constraints provided fewer and less consistent benefits. In an experiment designed to reduce the extent to which background knowledge could be exploited by the LLMs, LLM-based approaches still outperformed the baselines.
\end{enumerate}

\begin{credits}
\subsubsection{\ackname} 
Hira Beril Kucuk’s PhD research is supported by the Ministry of National Education of the Republic of Türkiye, and sincere thanks are extended for their generous support.

\subsubsection{\discintname}
The authors declare that they have no competing interests.
\end{credits}

\bibliographystyle{splncs04}
\bibliography{mybibliography}

@InProceedings{10.1007/978-3-032-02215-8_2,
author="Wrembel, Robert",
editor="Leung, Carson K.
and Dign{\"o}s, Anton
and Kotsis, Gabriele
and Tjoa, A. Min
and Khalil, Ismail",
title="Data Integration in the AI Era: Research Trends and Still Open Issues",
booktitle="Big Data Analytics and Knowledge Discovery",
year="2026",
pages="21--36",
isbn="978-3-032-02215-8"
}

@inproceedings{10.1145/2983323.2983767,
author = {Wang, Xianzhi and Sheng, Quan Z. and Yao, Lina and Li, Xue and Fang, Xiu Susie and Xu, Xiaofei and Benatallah, Boualem},
title = {Empowering Truth Discovery with Multi-Truth Prediction},
year = {2016},
isbn = {9781450340731},
booktitle = {Proc. 25th ACM CIKM},
pages = {881–890},
numpages = {10},
location = {Indianapolis, Indiana, USA},
series = {CIKM '16}
}

@article{Ji2025TableResolution,
    title = {Table integration in data lakes unleashed: pairwise integrability judgment, integrable set discovery, and multi-tuple conflict resolution},
    year = {2025},
    journal = {The VLDB Journal},
    author = {Daomin Ji and Hui Luo and Zhifeng Bao and J Shane Culpepper},
    pages = {36},
    volume = {34}
}

@inproceedings{QianHZHMWWSLDZ24,
  author       = {Yichen Qian and
                  others},
  title        = {UniDM: {A} Unified Framework for Data Manipulation with Large Language
                  Models},
  booktitle    = {Proceedings of the Seventh Annual Conference on Machine Learning and
                  Systems, MLSys 2024},
  year         = {2024},
  bibsource    = {dblp computer science bibliography, https://dblp.org}
}

@article{LiuPSWF25,
  author       = {Yurong Liu and
                  Eduardo Pe{\~{n}}a and
                  A{\'{e}}cio S. R. Santos and
                  Eden Wu and
                  Juliana Freire},
  title        = {Magneto: Combining Small and Large Language Models for Schema Matching},
  journal      = {Proc. {VLDB} Endow.},
  volume       = {18},
  number       = {8},
  pages        = {2681--2694},
  year         = {2025},
  doi          = {10.14778/3742728.3742757},
  bibsource    = {dblp computer science bibliography, https://dblp.org}
}

@inproceedings{DBLP:conf/esws/WuCP25,
  author       = {Zhenyu Wu and
                  Jiaoyan Chen and
                  Norman W. Paton},
  title        = {Taxonomy Inference for Tabular Data Using Large Language Models},
  booktitle    = {The Semantic Web - 22nd European Semantic Web Conference, {ESWC}, Part {I}},
  pages        = {403--422},
  year         = {2025},
  doi          = {10.1007/978-3-031-94575-5\_22},
  bibsource    = {dblp computer science bibliography, https://dblp.org}
}

@article{NobariR25,
  author       = {Arash Dargahi Nobari and
                  Davood Rafiei},
  title        = {TabulaX: Leveraging Large Language Models for Multi-Class Table Transformations},
  journal      = {Proc. {VLDB} Endow.},
  volume       = {18},
  number       = {11},
  pages        = {3826--3839},
  year         = {2025},
  doi          = {10.14778/3749646.3749657},
  bibsource    = {dblp computer science bibliography, https://dblp.org}
}

@article{ZeakisPSK25,
  author       = {Alexandros Zeakis and
                  George Papadakis and
                  Dimitrios Skoutas and
                  Manolis Koubarakis},
  title        = {An in-depth analysis of pre-trained embeddings for entity resolution},
  journal      = {{VLDB} J.},
  volume       = {34},
  number       = {1},
  pages        = {5},
  year         = {2025},
  doi          = {10.1007/S00778-024-00879-4},
  bibsource    = {dblp computer science bibliography, https://dblp.org}
}

@article{Dong0NEO23,
  author       = {Yuyang Dong and
                  Chuan Xiao and
                  Takuma Nozawa and
                  Masafumi Enomoto and
                  Masafumi Oyamada},
  title        = {DeepJoin: Joinable Table Discovery with Pre-trained Language Models},
  journal      = {Proc. {VLDB} Endow.},
  volume       = {16},
  number       = {10},
  pages        = {2458--2470},
  year         = {2023},
  doi          = {10.14778/3603581.3603587},
  bibsource    = {dblp computer science bibliography, https://dblp.org}
}

@article{steiner2026automaticendtoenddataintegration,
  author       = {Aaron Steiner and
                  Christian Bizer},
  title        = {Automatic End-to-End Data Integration using Large Language Models},
  journal      = {CoRR},
  volume       = {abs/2603.10547},
  year         = {2026},
  doi          = {10.48550/ARXIV.2603.10547},
  eprinttype   = {arXiv},
  eprint       = {2603.10547},
  bibsource    = {dblp computer science bibliography, https://dblp.org}
}

@article{10.1145/2897350.2897352,
author = {Li, Yaliang and Gao, Jing and Meng, Chuishi and Li, Qi and Su, Lu and Zhao, Bo and Fan, Wei and Han, Jiawei},
title = {A Survey on Truth Discovery},
year = {2016},
issue_date = {December 2015},
publisher = {Association for Computing Machinery},
address = {New York, NY, USA},
volume = {17},
number = {2},
issn = {1931-0145},
journal = {SIGKDD Explor. Newsl.},
month = feb,
pages = {1–16},
numpages = {16}
}

@misc{lunadong_datasets2026,
  author       = {Dong, Xin Luna},
  title        = {Fusion Datasets},
  year         = {2026},
  howpublished = {\url{https://lunadong.com/fusiondatasets}},
  note         = {Accessed: 2026-03-30}
}

@misc{heathersherry2026,
  title        = {heathersherry.github.io Personal Website},
  year         = {2026},
  howpublished = {\url{https://heathersherry.github.io/}},
  note         = {Accessed: 2026-03-30}
}

@article{LiTruthSolved,
author = {Li, Xian and Dong, Xin Luna and Lyons, Kenneth and Meng, Weiyi and Srivastava, Divesh},
title = {Truth finding on the deep web: is the problem solved?},
year = {2012},
issue_date = {December 2012},
publisher = {VLDB Endowment},
volume = {6},
number = {2},
issn = {2150-8097},
journal = {Proc. VLDB Endow.},
month = dec,
pages = {97–108},
numpages = {12}
}

@article{BodensohnBVSB25,
  author       = {Jan{-}Micha Bodensohn and
                  Ulf Brackmann and
                  Liane Vogel and
                  Anupam Sanghi and
                  Carsten Binnig},
  title        = {Unveiling Challenges for LLMs in Enterprise Data Engineering},
  journal      = {Proc. {VLDB} Endow.},
  volume       = {19},
  number       = {2},
  pages        = {196--209},
  year         = {2025},
  url          = {https://www.vldb.org/pvldb/vol19/p196-bodensohn.pdf},
  bibsource    = {dblp computer science bibliography, https://dblp.org}
}

@article{Dong2009IntegratingDependence,
author = {Dong, Xin Luna and Berti-Equille, Laure and Srivastava, Divesh},
title = {Integrating conflicting data: the role of source dependence},
year = {2009},
issue_date = {August 2009},
publisher = {VLDB Endowment},
volume = {2},
number = {1},
issn = {2150-8097},
journal = {Proc. VLDB Endow.},
month = aug,
pages = {550–561},
numpages = {12}
}

@article{DBLP:journals/tbd/WangZSLSCZYG25,
  author       = {Shuang Wang and
                  He Zhang and
                  Quan Z. Sheng and
                  Xiaoping Li and
                  Zhu Sun and
                  Taotao Cai and
                  Wei Emma Zhang and
                  Jian Yang and
                  Qing Gao},
  title        = {A Survey on Truth Discovery: Concepts, Methods, Applications, and
                  Opportunities},
  journal      = {{IEEE} Trans. Big Data},
  volume       = {11},
  number       = {2},
  pages        = {314--332},
  year         = {2025},
  bibsource    = {dblp computer science bibliography, https://dblp.org}
}

@article{PvldbReferenceFormatXueling2018Domain-AwareSources,
author = {Lin, Xueling and Chen, Lei},
title = {Domain-aware multi-truth discovery from conflicting sources},
year = {2018},
issue_date = {January 2018},
publisher = {VLDB Endowment},
volume = {11},
number = {5},
issn = {2150-8097},
journal = {Proc. VLDB Endow.},
month = jan,
pages = {635–647},
numpages = {13}
}

@article{Zhao2150AIntegration,
author = {Zhao, Bo and Rubinstein, Benjamin I. P. and Gemmell, Jim and Han, Jiawei},
title = {A Bayesian approach to discovering truth from conflicting sources for data integration},
year = {2012},
issue_date = {February 2012},
publisher = {VLDB Endowment},
volume = {5},
number = {6},
issn = {2150-8097},
journal = {Proc. VLDB Endow.},
month = feb,
pages = {550–561},
numpages = {12}
}

@article{Bleiholder2009DataFusion,
    title = {{Data Fusion}},
    year = {2009},
    journal = {ACM Computing Surveys},
    author = {Bleiholder, Jens and Naumann, Felix},
    number = {1},
    month = {1},
    pages = {1--41},
    volume = {41},
    issn = {15577341}
}

@article{DongDataIntegration,
author = {Dong, Xin Luna and Naumann, Felix},
title = {Data fusion: resolving data conflicts for integration},
year = {2009},
issue_date = {August 2009},
publisher = {VLDB Endowment},
volume = {2},
number = {2},
issn = {2150-8097},
journal = {Proc. VLDB Endow.},
month = aug,
pages = {1654–1655},
numpages = {2}
}

@article{Tian2025DataChallenges,
    title = {{Data Integration and Storage Strategies in Heterogeneous Analytical Systems: Architectures, Methods, and Interoperability Challenges}},
    year = {2025},
    journal = {Information 2025, Vol. 16, Page 932},
    author = {Tian, Yuan and Sun, Le and Potdar, Vidyasagar and Song, Biao and Koukaras, Paraskevas},
    number = {11},
    month = {10},
    pages = {932},
    volume = {16},
    publisher = {Multidisciplinary Digital Publishing Institute},
    issn = {2078-2489}
}

@article{Mandreoli2019DealingAlgorithms,
    title = {{Dealing With Data Heterogeneity in a Data Fusion Perspective: Models, Methodologies, and Algorithms}},
    year = {2019},
    journal = {Data Handling in Science and Technology},
    author = {Mandreoli, Federica and Montangero, Manuela},
    month = {1},
    pages = {235--270},
    volume = {31},
    publisher = {Elsevier},
    issn = {0922-3487}
}

@article{Putrama2024HeterogeneousOpportunities,
    title = {{Heterogeneous data integration: Challenges and opportunities}},
    year = {2024},
    journal = {Data in Brief},
    author = {Putrama, I. Made and Martinek, Péter},
    month = {10},
    pages = {110853},
    volume = {56},
    publisher = {Elsevier},
    issn = {2352-3409}
}

@article{Li2014ResolvingEstimation,
    title = {{Resolving conflicts in heterogeneous data by truth discovery and source reliability estimation}},
    year = {2014},
    journal = {Proc. ACM SIGMOD},
    author = {Li, Qi and Li, Yaliang and Gao, Jing and Zhao, Bo and Fan, Wei and Han, Jiawei},
    pages = {1187--1198},
    publisher = {Association for Computing Machinery},
    isbn = {9781450323765},
    issn = {07308078}
}

@article{Dong2150LessIntegration,
author = {Dong, Xin Luna and Saha, Barna and Srivastava, Divesh},
title = {Less is more: selecting sources wisely for integration},
year = {2012},
issue_date = {December 2012},
publisher = {VLDB Endowment},
volume = {6},
number = {2},
issn = {2150-8097},
journal = {Proc. VLDB Endow.},
month = dec,
pages = {37–48},
numpages = {12}
}

@inproceedings{Yin2007Truth,
author = {Yin, Xiaoxin and Han, Jiawei and Yu, Philip S.},
title = {Truth discovery with multiple conflicting information providers on the web},
year = {2007},
isbn = {9781595936097},
publisher = {ACM},
booktitle = {Proc. 13th ACM SIGKDD},
pages = {1048–1052},
numpages = {5},
location = {San Jose, California, USA},
series = {KDD '07}
}

@article{wei2022chain,
  title={Chain-of-thought prompting elicits reasoning in large language models},
  author={Wei, Jason and Wang, Xuezhi and Schuurmans, Dale and Bosma, Maarten and Xia, Fei and Chi, Ed and Le, Quoc V and Zhou, Denny and others},
  journal={Advances in neural information processing systems},
  volume={35},
  pages={24824--24837},
  year={2022}
}
\end{document}